\documentclass[aps,prl,twocolumn,superscriptaddress]{revtex4}
\usepackage{amsfonts}
\usepackage{amsmath}
\usepackage{amssymb}
\usepackage{graphicx}

\begin{document}

\title{A toy model mimicking cage effect, structural fluctuations and kinetic constraints in supercooled liquids.}


\author{V. Teboul}
\email{victor.teboul@univ-angers.fr}
\affiliation{ Laboratoire de Photonique d'Angers EA 4464, Universit\' e d'Angers, Physics Department,  2 Bd Lavoisier, 49045 Angers, France}

\keywords{dynamic heterogeneity,glass-transition}
\pacs{64.70.Q-,64.70.pm}

\begin{abstract}

The cage effect is widely accepted as the basic microscopic mechanism underlying the physics of supercooled liquids
 in contrast with usual liquids which are governed by molecular interactions only. 
In this work we implement a new toy model coined to reproduce the cage effect with variants including structural fluctuations and kinetic constraints. 
We use this new model to investigate which glass-transition features are directly due to the cage effect and which are due to more complex mechanisms.

\end{abstract}

\maketitle
To shed some light into the physics of the glass transition, we need toy models sufficiently simple to understand the phenomena but sophisticated enough to contain the whole set of relevant physical mechanisms.
Spin glasses\cite{kc1,kc2,kc3,kc4} are such models that have leaded in the past to a better understanding of the physics at work behind the glass-transition phenomena. 
However models in between spin glasses and real liquids are still lacking. 
The cage effect\cite{cage}, i.e. the hindrance of motion due to the presence of the first neighbors, is widely accepted as the first step microscopic mechanism underlying the physics of supercooled liquids in their approach to the glass-transition.
The model will thus have to use the cage effect as its fundamental microscopic step while being realistic enough to reproduce the main features of the glass-transition. 
In this work we present such a model and use it to investigate which effects in the glass-transition phenomenology\cite{gt1,rfot1} are directly due to the caging and which effects are due to more complex mechanisms. 
Our toy model consists in molecules moving inside a maze with doors that open and close randomly. The door size is in that model the parameter that replaces the temperature in real systems. The underlying idea is that as the temperature decreases the easily accessible passages in the energy landscape decrease in size, an effect modelled here with a decreasing door size, but in a more general picture the door surface is simply a parameter that sets in the model the probability to escape the cage after each collision.  
We observe with that simple model the appearance of most of the specific phenomena of the glass transition.
Then to test more complex mechanisms we add structural fluctuations and kinetic constraints\cite{kc1,kc2,kc3,kc4} in the model.
We find that if dynamic heterogeneities (DHs) \cite{dh-1,dh0,dh1,dh3,dh5,dh8,dh9,tc} are already present in the simple cage model, kinetic constraints are necessary to observe a variation of these cooperative motions when the relaxation times increase. 
Note that using modern computers our very simple toy model permits to access very large time scales ($>$ seconds) and hence makes possible the simulation within the model of the glass transition temperature phenomenology.

The model consists in a $70$ \AA\ wide simulation box with periodic limit conditions.  We then divide that box in $125$ ($5$ x $5$ x $5$) smaller cubic boxes $14$ \AA\ wide each (the surface of a wall is $S_{0}=196$\AA$^{2}$). Then in the center of each small box wall we manage a squared shape door that may be opened or closed. We use the model with ever a very low density ($10$ molecules) or a small density ($1250$ molecules i.e. an average of $10$ molecules in a box).   The small box do have reflecting limit conditions so that a molecule will bounce inside the box until it finds an opened door in its trajectory. The doors are set opened or closed with a random number generator for a period $\tau_{rand}=100 ps$.  
Depending on the model variant, the molecules constituted of two atoms ($i=1, 2$) do interact (or not) with the Lennard-Jones potential  
$V_{ij}=4\epsilon_{ij}((\sigma_{ij}/r)^{12} -(\sigma_{ij}/r)^{6})$ with the parameters: $\epsilon_{11}= \epsilon_{12}=0.5 KJ/mol$, $\epsilon_{22}= 0.4 KJ/mol$, $\sigma_{11}= \sigma_{12}=3.45$\AA\ and $\sigma_{22}=3.28$\AA$ $.
We use the mass of Argon for each atom of the molecule that we rigidly bonded fixing the interatomic distance to $d=1.73 $\AA$ $.  
 We include these interactions to increase randomness and eliminate possible trapping in closed trajectories inside the boxes, however at the temperature of study ($T=300K$) the corresponding liquid is above its melting temperature, insuring that the observed effects in the model are due to the cages only, and removing totally the interaction has in most cases only small effects. 
We study several variants of that model. In a first variant the size of the doors is distributed at random and modified each $100 ps$. That variant leads to a continuous distribution of environments that results randomly to soft zones (large doors or most doors opened) and hard zones (small doors or most doors closed). We use that variant to model the local structure fluctuations. 
In a second variant we introduce the following kinetic constraint: The door is opened or closed depending on the total number of molecules in the two boxes between which the door is opened. We tested various conditions for the door opening and use the following one here: below a threshold number of molecules the door is opened and otherwise the door is closed. Our purpose here is both to mimic the kinetic constraint used in spin glasses models\cite{kc1,kc2,kc3}, and to take into account the retroaction induced by local density fluctuations in real liquids.

We show in Figure 1 the mean square displacements ($<r^{2}(t)>$, MSD) for the simplest of our toy models. 
We observe the three characteristic time scales of supercooled liquids, namely the ballistic time scale, then the plateau due to the trapping of the molecule inside the cage and the diffusive time scale when the molecule finally escapes the cage. As the doors surfaces decrease the plateau increases leading to a decrease of the diffusion coefficient. The ballistic part of the MSD is here entirely defined by the temperature T ($<r^{2}(t)>=3.k_{B}.T.t^{2}/m$) while the plateau and diffusive part depend on the door surfaces only. 

\includegraphics[scale=0.32]{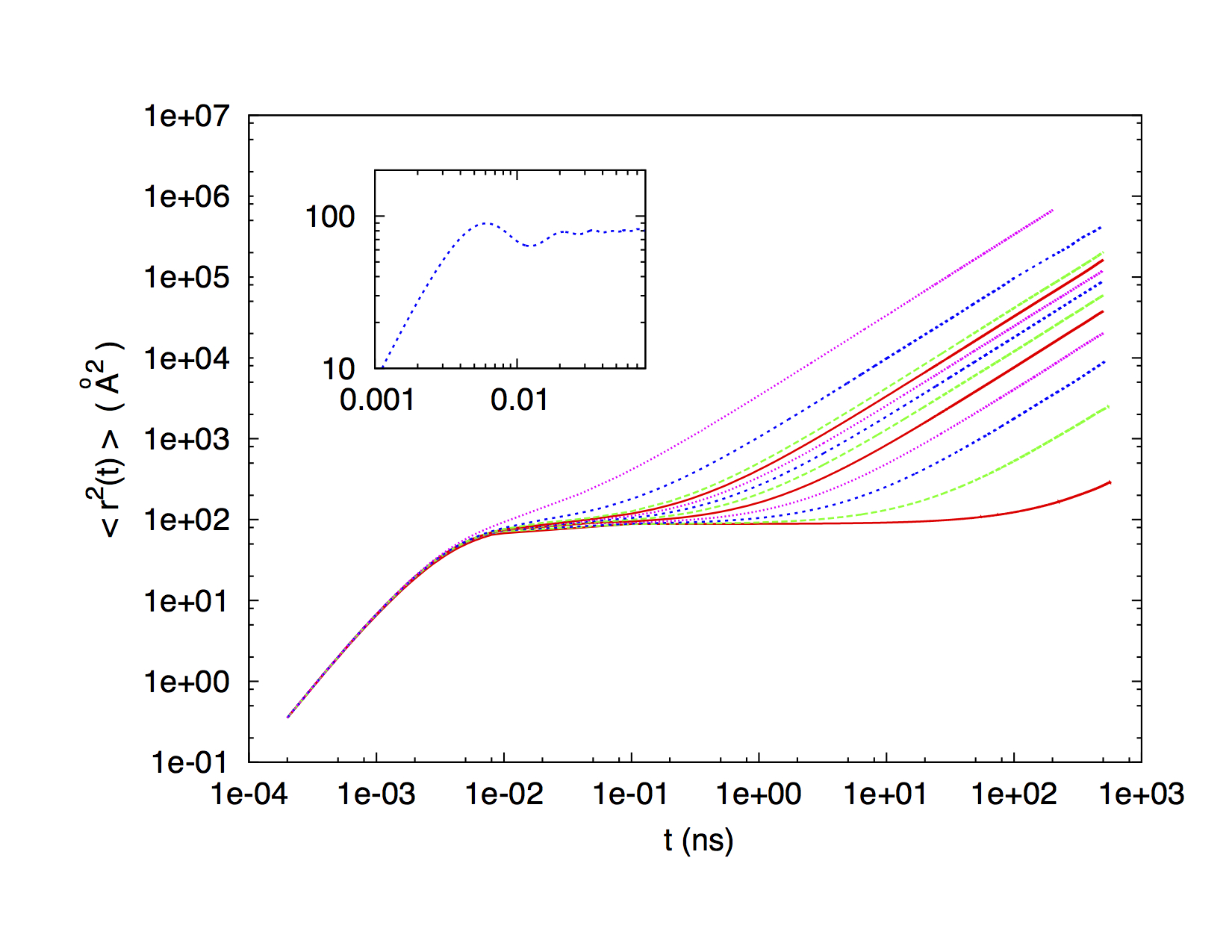}

{\em \footnotesize  FIG.1. (color online) Mean square displacements for various average door surfaces for N=1250 molecules. From bottom to top $S= 0.04, 0.16, 0.36, 0.64, 1.$$, 1.44, 1.96, 2.56, 3.24, 4.0, 9.0,$ and $36 $\AA$^{2}$. The average surface of the door is the parameter that represents the temperature in the model. Inset: Same curve and axis but for N=10 molecules and $S=0.36 $\AA$^{2}$. The oscillations that are washed out for larger numbers of molecules are here clearly visibles.\\}

When small numbers of molecules are considered we observe oscillations in the MSD in the approach of the plateau time scale (see Fig. 1 inset). These oscillations arise here due to the bouncing of molecules on the cages walls and are also present in simulations of real glass-formers where they are most often related to the Boson peak unsolved phenomenon. Note that the mean square displacements in Figure 1 resembles so much the MSD of real supercooled liquids that it is not possible to tell from the Figure that this is not the MSD of a real liquid. For the larger doors used in the Figure ($S=36$ \AA$^{2}$), the plateau disappears, reproducing the behavior of liquids above their melting temperature.

Figure 2 shows the self Van Hove correlation function $G_{S}(r,t)$ that represents the probability for a molecule to be after a time lapse $t$, at a distance $r$ from its initial position, while the inset in Figure 2 shows the mean square displacement together with the non Gaussian parameter $\alpha_{2}(t)=\frac{3 <r^{4}(t)>} {5<r^{2}(t)>^{2}}    -1$  that measures the deviation of the Van Hove from the Gaussian shape predicted by Brownian motion. The inset shows that $\alpha_{2}(t)$ is maximum at the plateau ending of $<r^{2}(t)>$, a result in amazing agreement with real supercooled liquids behavior.
Interestingly the Van Hove correlation functions in Figure 2 also display a tail similar to the tails observed in supercooled liquids and that are fingerprints of the dynamic heterogeneities. The tail is due in our model to the contribution of the molecules that find the way out the cage through the doors. 
The tail is thus induced by the cage escaping process, a result that explains the $\alpha_{2}(t)$ location at the end of the plateau of $<r^{2}(t)>$.
Note that in the case of a density fluctuation between two nearby cages (i.e. boxes) non mobile molecules are structurally more correlated than the mean in the model as they remain in the same cage, while most mobile molecules are also structurally correlated as they have some chance to have moved in the same nearby cage.  As a result the model spontaneously creates dynamic heterogeneities. 

\includegraphics[scale=0.32]{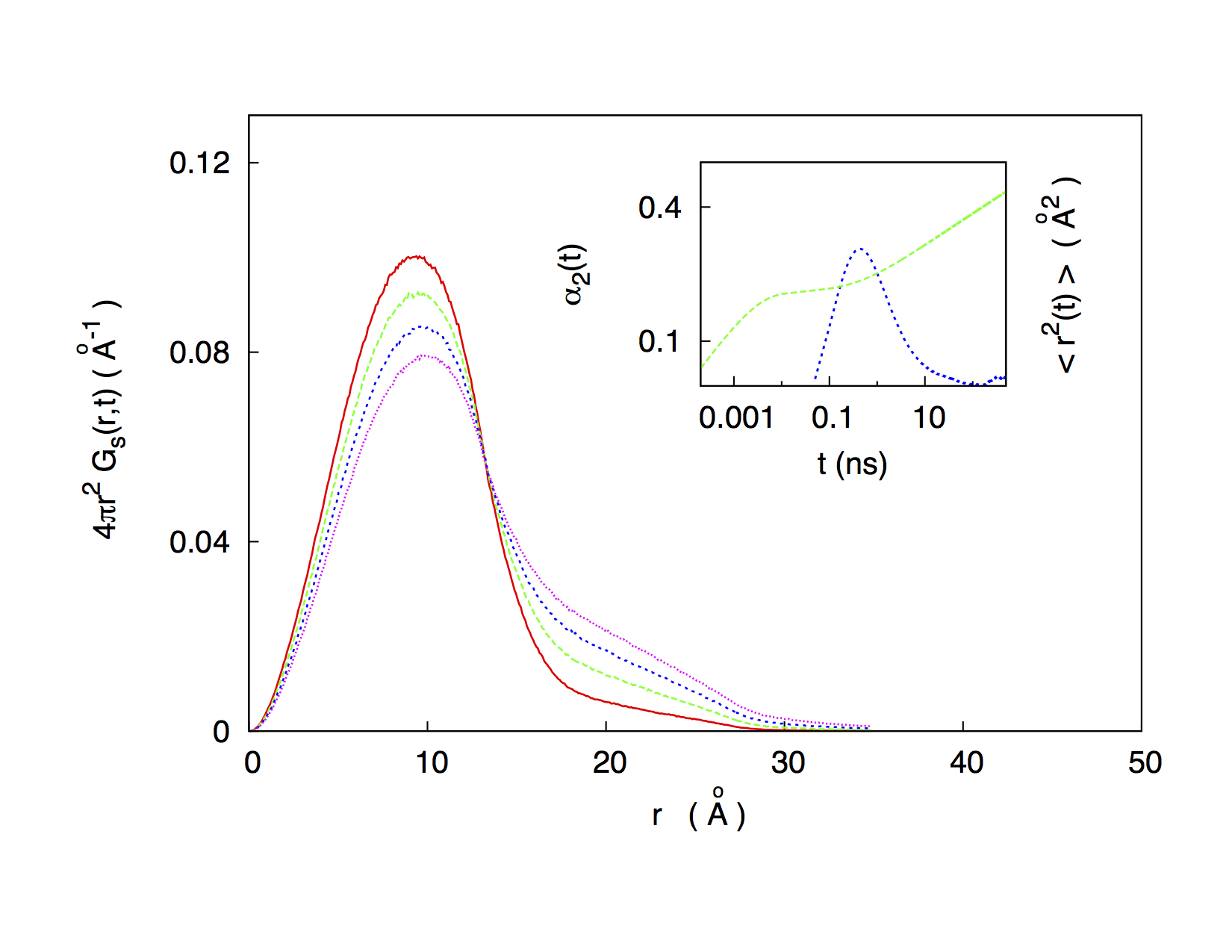}

{\em \footnotesize  FIG.2. (color online) Van Hove self correlation function for various time lapses. From top to bottom at $r=10$ \AA: $t= 50, 100, 150$ and $200 ps$. $S=2.56$ \AA$^{2}$. We see the appearance of a tail that is present in real glass-formers where it is usually related to the presence of dynamic heterogeneities as the tail shows the presence of  anomalously mobile molecules. Here the tail is due to the mobility difference between the few molecules moving outside the cage and the rest. That effect is clearly present in real glass-formers but may be enhanced by other mechanisms. Inset: Mean square displacement in logarithmic scale ($<r^{2}(t)>$, green dashed line, \AA$^{2}$)  and non Gaussian parameter $\alpha_{2}(t)$ in linear scale (blue dashed line). Conditions are the same: $S=2.56$ \AA$^{2}$. The plot shows that $\alpha_{2}(t)$ is maximum at the plateau ending of $<r^{2}(t)>$, a result in agreement with real supercooled liquids behavior.\\}

However these heterogeneities do not evolve with the doors size while in real glass formers the DHs increase when the temperature decreases. 
Indeed investigating the non Gaussian parameter $\alpha_{2}(t)$ evolution with $S$, we find that $\alpha_{2}(t)$ is shifted to larger times when $S$ decreases but otherwise is roughly constant. For the model with 50 percents opened doors for example we find $\alpha_{2}(t^{*})=0.30  \pm  0.01$  while $t^{*}$ increased from $0.28 ns$ for $S=4$ \AA$^{2}$ to $289 ns$ for $S=0.04$ \AA$^{2}$. 
To confirm the DHs evolution with $S$ that we observe with $\alpha_{2}$ we also studied the dynamic susceptibility $\chi _{4}$ evolution \cite{rfot1,dh1}:
 $\chi _{4}(a,t)=\frac{\beta V}{N^{2}}\left( \left\langle
C_{a}(t)^{2}\right\rangle -\left\langle C_{a}(t)\right\rangle ^{2}\right)$ with $C_{a}(t)={\sum_{i=1}^{N}{w_{a}}}\left( \left\vert {{{\mathbf{r}}}}_{i}(t)-{{{\mathbf{r}}}}_{i}(0)\right\vert \right)$
In that formula $V$ is the volume of the simulation box, $N$ the number of molecules, and $\beta =(k_{B}T)^{-1}$. The symbol $w_{a}$ stands for a discrete mobility  window function, $w_{a}(r)$, taking the values $w_{a}(r)=1$ for $r>a$ and zero otherwise. 
We chose in this work $a=14$ \AA\ that is the size of the cage in our model, a value that insures us that we are probing diffusive motions fluctuations. 
Finally we find the same behavior (i.e. roughly constant and shifted in times) for the dynamic susceptibility than for the non Gaussian parameter, confirming that the DHs stay constant in that model.
To conclude that first part, the pure cage effect model suggests a decoupling between the dynamic heterogeneities and the structural relaxation evolution, as in the model the slowing down takes place as a result of the cages closing while the mobility aggregation is not impacted.
Note that a limitation of the model for the facilitation mechanism is the absence of connections between the number of molecules in a cage and the doors opening. The molecules motion outside the cage will then not trigger the motions of other molecules in the model. We will thus include now this mechanism for the purpose of better understanding the origins of the DHs increase when the relaxation time increases.
We tested various retroactions (kinetic constraints) 
the results being qualitatively similar, we chose here one of the simplest.
Our constraint consists in opening the door only if the sum of the number of molecules in the two boxes connected by that door is below a threshold value $2.K$.
As the mean number of molecules in a box $<n>=10$, we expect the constraint to act efficiently for $K<10$ and to be weak above. 
To simplify the interpretation we remove the intermolecular interactions in that model.

We show in Figure 3 the evolution of the dynamic susceptibility $\chi_{4}$ and non Gaussian parameter $\alpha_{2}$ with the threshold $K$ of our kinetic constraint. 
Figure 3 shows that when the kinetic constraint is made stronger (i.e. when $K$ decreases) the dynamic susceptibility $\chi_{4}$ and the NGP $\alpha_{2}$ increase, showing that the dynamic heterogeneities increase with the constraint while the characteristic times $t^{*}$  and $t^{+}$, defined as $\alpha_{2}(t^{*})=max(\alpha_{2}(t))$ and $\chi_{4}(t^{+})=max(\chi_{4}(t))$, increase slightly. 
As a result the phenomenology of the glass-transition is reproduced with our model if an increase of the kinetic constraint is associated to the decrease of the probability $p$ to escape the cage (i.e. a decrease of the door surface).
That picture is correct for example if the kinetic constraint (here a condition in the very local density) is one of the origins of the cage escape process.
In that picture as the temperature decreases the energy required to open the cage is less often encountered in thermal fluctuations and the conditions on density fluctuations become thus more stringent.

\includegraphics[scale=0.32]{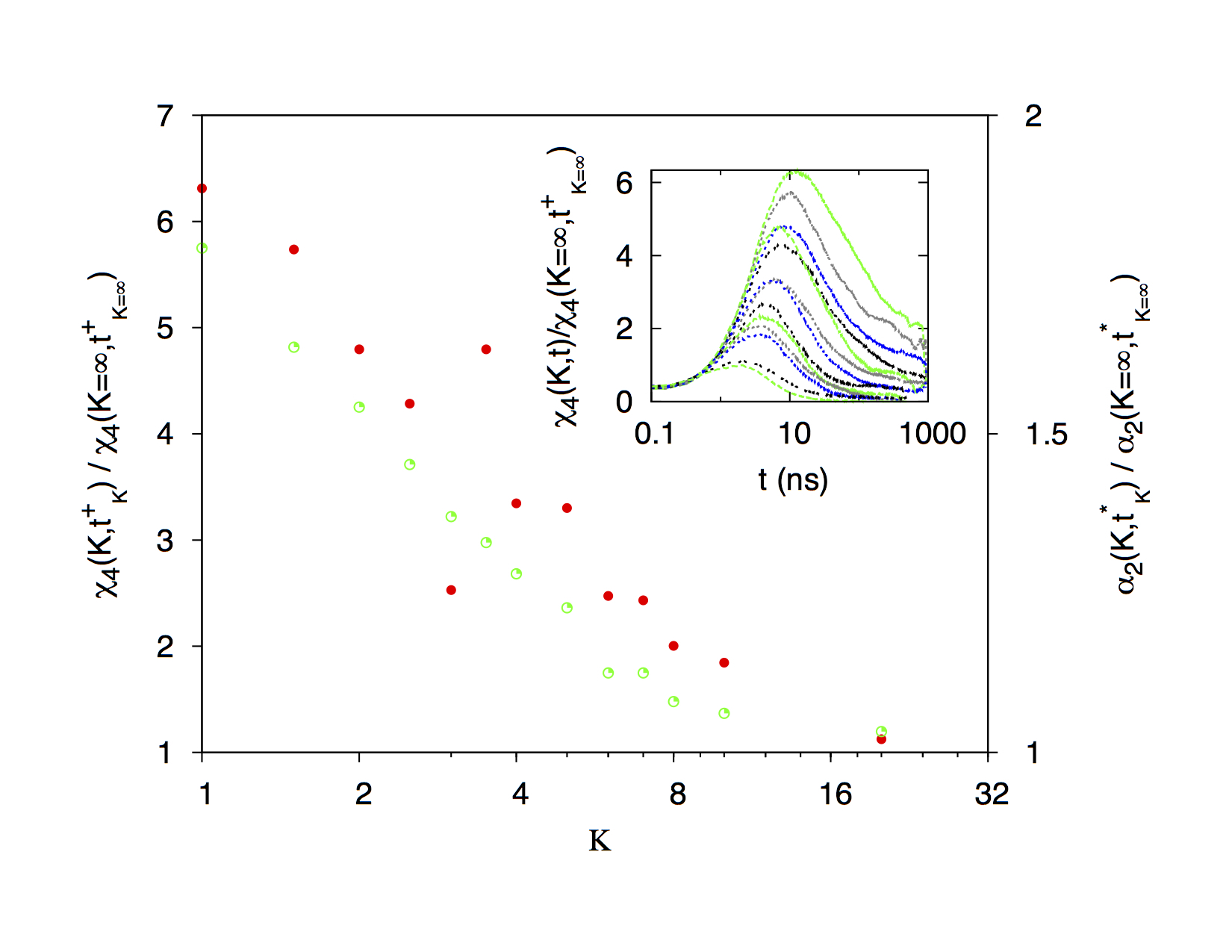}

{\em \footnotesize  FIG.3. (color online)  Maximum values of the non Gaussian parameter $\alpha_{2}$ (open green circles) and of the dynamic susceptibility $\chi_{4}$ ( solid red circles) versus kinetic constraint threshold $K$ (in logarithmic scale) for $S=1$\AA$^{2}$. An increase in $K$ in the model decreases the constraint. The $\chi_{4}$ or $\alpha_{2}$  values are rescaled by the maximum value of the dynamic susceptibility respectively non Gaussian parameter with $K=\infty$ corresponding to a released kinetic constraint. We find a large fluctuation of the susceptibility around $K=3$ otherwise $\chi_{4}$ increases monotonously when $K$ decreases. Inset:  Dynamic susceptibility $\chi_{4}(t)$  for various constraints $K$. From top to bottom $K= 1,1.5,2,2.5,3.5,4,5,6,7,8,10,20,\infty.$ The curves are rescaled by the same maximum value of the dynamic susceptibility without constraint.\\}

To test the effect of structural fluctuations that is another possible cause for DHs, we modify now the first model by replacing the fixed door surfaces with randomly chosen surfaces for each door.
Including that fluctuations results in an increase of the DHs however much slighter than the increase observed when we introduced the kinetic constraints.
Again the DHs do not increase within that model when the mean door surface decreases. 
The diffusion is also only slightly modified in that model as shown in Figure 4 that displays the evolution of the diffusion coefficient $D$ with the doors mean surface $S$ (for the two densities investigated and various models). The green dashed line in the Figure results from
the assumption that for each collision with a wall the probability $p$ to escape the cage is the ratio of the door to the wall surfaces $p=S/S_{0}$, leading to 
$D=(S/S_{0}).(1/6.t_{0})$ where $t_{0}$ is the time between two collisions i.e. the time to cross the box ballistically.  
Figure 4 shows that for large doors, the diffusion coefficient follows the probabilistic law described above, then deviates from that law when $S$ becomes to be small tending to a law of the form $D=\alpha (S/S_{0})^{1.5}$.
We interpret that deviation as arising from percolation phenomena in our maze that are not included in the simple linear probabilistic law.
Interestingly that deviation is reminiscent of the non-Arrhenius behavior with temperature of fragile glass-formers.
Indead, relating the probability $p$ to escape the cage to an activation energy  
$p=(S/S_{0}) \sim e^{-E_{a}/k_{B}T}$, (in the picture of an energy barrier to overpass to open the cage) the deviation of Figure 4 from the linear law leads to a non-Arrhenius behavior of the diffusion with temperature, the activation energy increasing by a factor $1.5$.
Figure 4 also shows very similar diffusive comportment for the various models. That result suggests that if diffusion depends strongly on the cage opening probability (i.e. here on the mean door size), it depends more weakly on structural fluctuations (i.e. fluctuations of the door sizes),  a result reminiscent of the recent finding\cite{t} that similar potentials with different attractive parts (thus different cage breaking probabilities) lead to different diffusive comportment while structures are similar.

\includegraphics[scale=0.18]{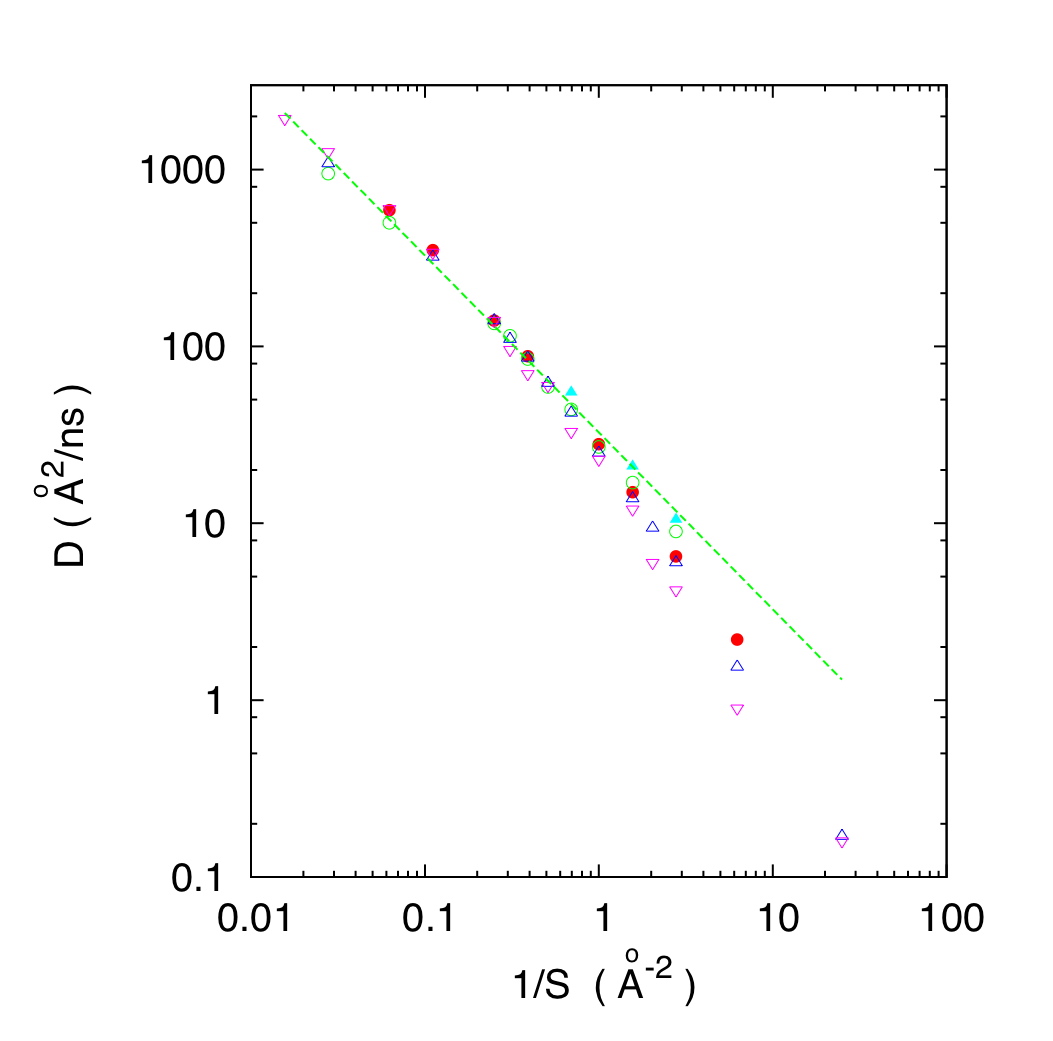}

{\em \footnotesize  FIG.4. (color online) Diffusion coefficient $D$ versus doors average surfaces $S$ for various models. $D$ is here corrected from the percentage of doors opened. The dashed line is the simple law $D=(S/S_{0}).(1/6.t_{0})$ that results from the assumption that the probability to escape the cage is proportional to the surface of the door $S$ divided by the surface of the wall $S_{0}$, where $t_{0}$ is the time necessary to cross ballistically the cage at the temperature of study. There is 1050 molecules in the maze otherwise it is specified. The models are as follows: Solid (red) circles: the 6 doors are opened for every boxes. Open (blue) triangles (1050 molecules in the maze) and open (pink) inverted triangles (10 molecules only): We chose randomly 50 percents of the doors to be closed (hence some boxes may be totally opened or totally closed, but that random choice is reinitialized periodically each 100 ps time lapse). Open (green) circles: 20 percents of the doors are opened randomly. Solid (blue) triangles: We use a continuous random distribution of surfaces for the doors.\\}

To summarize, we have implemented toy models, situated in between constrained spin glasses and real liquids, and intended to serve as paradigms for studying the glass-transition physical mechanisms. Our simplest model based on the cage effect, reproduces most of the features of the dynamics of supercooled liquids, namely the plateau and small oscillations in the density correlation functions, the three different time scales (ballistic, plateau and diffusive), the non-Gaussian behavior of the Van Hove correlation function with a maximum of the non-Gaussian parameter localized in time around the plateau regime ending, and the appearance of a tail in the self Van Hove correlation function.  We also observe a door size equivalent to the melting temperature, and the appearance of the cooperative effects called dynamic heterogeneities. But in that model the dynamic heterogeneities do not increase with the relaxation time (i.e. the decrease of the probability to escape the cage) and when we include structural fluctuations in the model, these results do not change drastically.
However dynamic heterogeneities increase significantly when we include kinetic constraints inside the model.
Thus our results suggest that the increase of the DHs when the temperature decreases is due to the hardening of kinetic constraints as opposed to a structural fluctuation origin.
Our results also suggest that the connection between the DHs and the dynamical slowing down is only indirect, as in our models the slowing down is induced by the cage effect while the DHs are induced by the kinetic constraints, however the possibility that in real liquids, kinetic constraints are 
together with thermal fluctuations at the origin of the cage breaking mechanism could explain why a decoupling between DHs and the dynamical slowing down is so difficult to achieve.

\newpage
\textbf{Figure captions}
\vskip 0.5cm

FIG.1. (color online) Mean square displacements for various average door surfaces for N=1250 molecules. From bottom to top $S= 0.04, 0.16, 0.36, 0.64, 1.$$, 1.44, 1.96, 2.56, 3.24, 4.0, 9.0,$ and $36 $\AA$^{2}$. The average surface of the door is the parameter that represents the temperature in the model. Inset: Same curve and axis but for N=10 molecules and $S=0.36 $\AA$^{2}$. The oscillations that are washed out for larger numbers of molecules are here clearly visibles.\\

FIG.2. (color online) Van Hove self correlation function for various time lapses. 
From top to bottom at $r=10$ \AA: $t= 50, 100, 150$ and $200 ps$.
$S=2.56$ \AA$^{2}$.
We see the appearance of a tail that is present in real glass-formers where it is usually related to the presence of dynamic heterogeneities as the tail shows the presence of  anomalously mobile molecules. Here the tail is due to the mobility difference between the few molecules moving outside the cage and the rest.
That effect is clearly present in real glass-formers but may be enhanced by other mechanisms.
Inset: Mean square displacement in logarithmic scale ($<r^{2}(t)>$, green dashed line, \AA$^{2}$)  and non Gaussian parameter $\alpha_{2}(t)$ in linear scale (blue dashed line). Conditions are the same: $S=2.56$ \AA$^{2}$. The plot shows that $\alpha_{2}(t)$ is maximum at the plateau ending of $<r^{2}(t)>$, a result in agreement with real supercooled liquids behavior.\\

FIG.3. (color online)  Maximum values of the non Gaussian parameter $\alpha_{2}$ (open green circles) and of the dynamic susceptibility $\chi_{4}$ ( solid red circles) versus kinetic constraint threshold $K$ (in logarithmic scale) for $S=1$\AA$^{2}$. An increase in $K$ in the model decreases the constraint. 
The $\chi_{4}$ or $\alpha_{2}$  values are rescaled by the maximum value of the dynamic susceptibility respectively non Gaussian parameter with $K=\infty$ corresponding to a released kinetic constraint. We find a large fluctuation of the susceptibility around $K=3$ otherwise $\chi_{4}$ increases monotonously when $K$ decreases.
Inset:  Dynamic susceptibility $\chi_{4}(t)$  for various constraints $K$. From top to bottom $K= 1,1.5,2,2.5,3.5,4,5,6,7,8,10,20,\infty.$
The curves are rescaled by the same maximum value of the dynamic susceptibility without constraint.\\

FIG.4. (color online) Diffusion coefficient $D$ versus doors average surfaces $S$ for various models. 
$D$ is here corrected from the percentage of doors opened.
The dashed line is the simple law $D=(S/S_{0}).(1/6.t_{0})$ that results from the assumption that the probability to escape the cage is proportional to the surface of the door $S$ divided by the surface of the wall $S_{0}$, where $t_{0}$ is the time necessary to cross ballistically the cage at the temperature of study.
There is 1050 molecules in the maze otherwise it is specified. 
The models are as follows:
Solid (red) circles: the 6 doors are opened for every boxes.
Open (blue) triangles (1050 molecules in the maze) and open (pink) inverted triangles (10 molecules only): We chose randomly 50 percents of the doors to be closed (hence some boxes may be totally opened or totally closed, but that random choice is reinitialized periodically each 100 ps time lapse).
Open (green) circles: 20 percents of the doors are opened randomly.
Solid (blue) triangles: We use a continuous random distribution of surfaces for the doors.\\


\begin{thebibliography}{99}

\bibitem{kc1} F. Ritort, P. Sollich, \newblock  {\em Adv. Phys.} {\bf 52}, 219 (2003)
\bibitem{kc2} G.H. Fredrickson, H.C. Andersen, \newblock  {\em Phys. Rev. Lett.} {\bf 53}, 1244 (1984)
\bibitem{kc3} J. Jackle, S. Eisinger, \newblock  {\em Z. Phys. B} {\bf 84}, 115 (1991)
\bibitem{kc4} W. Kob, H.C. Andersen, \newblock  {\em Phys. Rev. E} {\bf 48}, 4364 (1993)


\bibitem{cage} M. Goldstein, 
 \newblock \emph{J. Chem. Phys.} \textbf{51}, 3728 (1969)


\bibitem{gt1} K. Binder, W. Kob, 
\newblock  {\em  Glassy materials and disordered solids}, World Scientific, Singapore (2011)

\bibitem{rfot1} P.G. Wolynes, V. Lubchenko, 
\newblock  {\em  Structural Glasses and Supercooled Liquids}, Wiley, Hoboken, New Jersey (2012)



\bibitem{dh-1} W. Kob, C. Donati, S.J. Plimpton, P.H. Poole, S.C. Glotzer, \newblock  {\em Phys. Rev. Lett.} {\bf79}, 2827 (1997)
\bibitem{dh0} C. Donati, J. Douglas, W. Kob, et al,\newblock  {\em Phys. Rev. Lett.} {\bf 80}, 2338 (1998)

\bibitem{dh1} L. Berthier, G. Biroli, J.-P. Bouchaud, L. Cipelletti, W. Van Saarloos,
 \emph{Dynamical heterogeneities in glasses, colloids and granular
media} Oxford University Press, New York (2011).


\bibitem{dh3} D. Chandler, J.P. Garrahan,  \emph{Annu. Rev. Phys. Chem.} \textbf{61}, 191 (2010)


\bibitem{dh5} R. Yamamoto, A. Onuki, \newblock  {\em Phys. Rev. Lett.} {\bf81}, 4915 (1998)
\bibitem{dh8}  V. Teboul, M. Saiddine, J. M. Nunzi, 
\newblock  {{\em Phys. Rev. Lett. }} {\bf  103}, 265701 (2009)
\bibitem{dh9}  R.K. Darst, D.R. Reichman, G. Biroli,
\newblock  {{\em J. Chem. Phys.}} {\bf  132}, 044510 (2010)



\bibitem{tc} R. Candelier, A. Widmer-Cooper, J.K. Kummerfeld, O. Dauchot, G. Biroli, P. Harrowell, D.R. Reichman, \newblock  {\em Phys. Rev. Lett.} {\bf 105}, 135702 (2010) 


\bibitem{t} L. Berthier, G.Tarjus, \newblock  {\em Phys. Rev. Lett.} {\bf 103}, 170601 (2009) 




\end{thebibliography}
\end{document}